\documentclass{article}
\pdfoutput=1

\PassOptionsToPackage{numbers, compress}{natbib}

\usepackage[sglblindworkshop, final]{ai4nextg_neurips_2025}

\usepackage[utf8]{inputenc} 
\usepackage[T1]{fontenc}    
\usepackage{hyperref}       
\hypersetup{
    pdftitle={Masked Symbol Modeling for Demodulation of Oversampled Baseband Communication Signals in Impulsive Noise-Dominated Channels},
    pdfauthor={Oguz Bedir, Nurullah Sevim, Mostafa Ibrahim, Sabit Ekin},
    pdfsubject={NeurIPS 2025 Workshop: AI and ML for Next-Generation Wireless Communications and Networking (AI4NextG)},
    pdfkeywords={Transformer networks, self-supervised learning, oversampled baseband signals, pulse shaping, symbol demodulation, physical layer, impulsive noise, intelligent receivers}
}

\usepackage{url}            
\usepackage{booktabs}       
\usepackage{amsfonts}       
\usepackage{nicefrac}       
\usepackage{microtype}      
\usepackage{xcolor}         
\usepackage{acro}
\usepackage{subcaption}
\usepackage{amsmath, amssymb}
\usepackage{dsfont}
\usepackage{graphicx}

\title{Masked Symbol Modeling for Demodulation of Oversampled Baseband Communication Signals in Impulsive Noise-Dominated Channels}

\author{%
  Oguz~Bedir\thanks{These authors contributed equally to this work.} \\
  Electrical \& Computer Engineering\\
  Texas A\&M University\\
  College Station, TX 77843\\
  \texttt{oguzbedir@tamu.edu} \\
  \And
  Nurullah~Sevim\footnotemark[1] \\
  Electrical \& Computer Engineering\\
  Texas A\&M University\\
  College Station, TX 77843\\
  \texttt{nurullahsevim@tamu.edu} \\
  \And
  Mostafa~Ibrahim\\
  Engineering Technology \&\\ Industrial Distribution\\
  Texas A\&M University\\
  College Station, TX 77843\\
  \texttt{mostafa.ibrahim@tamu.edu} \\
  \And
  Sabit~Ekin\\
  Engineering Technology \& \\ Industrial Distribution, and \\
  Electrical \& Computer Engineering \\
  Texas A\&M University\\
  College Station, TX 77843\\
  \texttt{sabitekin@tamu.edu} \\
}
\DeclareAcronym{nlp}{
  short = NLP,
  long = natural language processing
}

\DeclareAcronym{isc}{
  short = ISC,
  long = inter-symbol contribution 
}

\DeclareAcronym{msm}{
  short = MSM,
  long = Masked Symbol Modeling
}

\DeclareAcronym{phy}{
  short = PHY,
  long = physical
}

\DeclareAcronym{id}{
    short = ID,
    long = identifier,
    short-plural-form = IDs,
    long-plural-form = identifiers
}

\DeclareAcronym{psk}{
  short = PSK,
  long = Phase-Shift Keying
}

\DeclareAcronym{bpsk}{
  short = BPSK,
  long = Binary Phase-Shift Keying
}

\DeclareAcronym{qam}{
  short = QAM,
  long = Quadrature Amplitude Modulation
}

\DeclareAcronym{nn}{
  short = NN,
  long = neural network
}

\DeclareAcronym{rf}{
  short = RF,
  long = Radio Frequency 
}

\DeclareAcronym{rc}{
  short = RC,
  long = raised cosine 
}

\DeclareAcronym{sps}{
  short = SPS,
  long = samples per symbol 
}

\DeclareAcronym{fir}{
  short = FIR,
  long = finite impulse response 
}

\DeclareAcronym{lsh}{
  short = LSH,
  long = locality-sensitive hashing 
}

\DeclareAcronym{ser}{
  short = SER,
  long = Symbol Error Rate 
}

\DeclareAcronym{snr}{
  short = SNR,
  long = Signal-to-Noise Ratio
}

\DeclareAcronym{pdf}{
  short = PDF,
  long = probability density function
}

\DeclareAcronym{llm}{
  short = LLM,
  long = large language model
}

\DeclareAcronym{ai}{
  short = AI,
  long = artificial intelligence
}

\DeclareAcronym{iid}{
  short = i.i.d,
  long = independent and identically distributed
}

\DeclareAcronym{kl}{
  short = KL,
  long = Kullback–Leibler divergence
}

\DeclareAcronym{bert}{
  short = BERT,
  long = Bidirectional Encoder Representations from Transformers
}
\begin{document}
\maketitle

\begin{abstract}
Recent breakthroughs in \acl{nlp} show that attention mechanism in Transformer networks, trained via masked-token prediction, enables models to capture the semantic context of the tokens and internalize the grammar of language. While the application of Transformers to communication systems is a burgeoning field, the notion of context within physical waveforms remains under‑explored. This paper addresses that gap by re-examining \ac{isc} caused by pulse-shaping overlap. Rather than treating \ac{isc} as a nuisance, we view it as a deterministic source of contextual information embedded in oversampled complex baseband signals. We propose \ac{msm}, a framework for the \ac{phy} layer inspired by Bidirectional Encoder Representations from Transformers methodology. In \ac{msm}, a subset of symbol-aligned samples is randomly masked, and a Transformer predicts the missing symbol \aclp{id} using the surrounding “in-between” samples. Through this objective, the model learns the latent syntax of complex baseband waveforms. We illustrate \ac{msm}’s potential by applying it to the task of demodulating signals corrupted by impulsive noise, where the model infers corrupted segments by leveraging the learned context. Our results suggest a path toward receivers that interpret, rather than merely detect communication signals, opening new avenues for context‑aware \ac{phy} layer design.
\end{abstract}
\acresetall

\section{Introduction}
\label{sec:intro}
The success of \acp{llm} in \ac{nlp} is largely attributed to the Transformer architecture \cite{vaswani:2017}. Its attention mechanism enables models to capture long-range dependencies and understand the semantic context in which words appear. This paradigm has fueled successful applications in diverse fields \cite{dosovitskiy:2021, senior:2020, chiu:2018}, including communication systems \cite{choukrun:2022, chen:2020, zecchin:2024, ott:2024}. However, while Transformers are increasingly used to solve complicated communication problems, the fundamental notion of context within physical waveforms remains largely unexplored. This paper aims to bridge that gap.

The overlap between adjacent pulses, inherent in pulse-shaping, embeds a predictable structure in the oversampled baseband signal. Each sample, therefore, contains information not only about its primary symbol but also about its neighbors, creating what we term \acp{isc}. In oversampled systems, \acp{isc} have typically been harnessed within equalization methods, but their broader potential remains under-explored. We reframe them instead as a deterministic source of information that a sophisticated model can exploit to learn contextual representations.

To exploit this structure, we introduce \ac{msm}, a self-supervised, \ac{bert}-style \cite{devlin:2019} framework designed for the \ac{phy} layer. \ac{msm} trains a Transformer to predict randomly masked symbols by analyzing the surrounding unmasked waveform samples. Through this objective, the model learns to internalize the underlying structure of pulse-shaped signals, moving beyond simple detection towards a more comprehensive interpretation of the waveform.

Our main contributions are as follows:
\begin{itemize}
    \item We introduce \ac{msm}, a novel framework for learning representations of oversampled complex-valued baseband signals by treating \ac{isc} as a source of contextual information.
    \item We demonstrate the practical utility of \ac{msm} by applying it to symbol prediction under impulsive noise (e.g., Middleton Class-A \cite{middleton:1977, berry:1981}), where the model leverages learned context to recover corrupted symbols.
\end{itemize}
Our results suggest a path toward receivers that interpret, rather than merely detect, communication signals, opening new avenues for designing intelligent and context-aware \ac{phy} layer.

The rest of the paper is organized as follows: Section~\ref{sec:preliminaries} introduces the signal and noise model used in this work. Section~\ref{sec:methodology} presents the proposed \ac{msm} framework and its architectural details. Section~\ref{sec:experimental-setup-results} describes the experimental setup and presents our results. Section~\ref{sec:conclusion} concludes the paper and discusses future directions.
\section{Preliminaries}
\label{sec:preliminaries}
\subsection{Digital Communication Basics}
\label{subsec:signal-model}
Let \(\{x[k]\} \in \mathbb{C}^{N}\) be a symbol sequence, drawn from a modulation set \(\mathcal{M}\), and \(g(t)\) the pulse shaping filter. The continuous-time baseband signal can be expressed as \cite{bjornson:2024}
\begin{equation}
    \label{eq:ct-clean-signal}
   s(t) = \sum_{k=0}^{N-1} x[k] \cdot g(t - kT),
\end{equation}
where \(T\) is the symbol duration. Sampling at \(L\) \ac{sps} produces the corresponding discrete-time baseband signal
\begin{equation}
    \label{eq:dt-clean-signal}
    s[n] = s(nT_{s}) = \sum_{k=0}^{N-1} x[k] \cdot g(nT_{s} - kT),
\end{equation}
where \(T_{s} = \frac{T}{L}\).
\subsection{Noise Model}
A widely used impulsive-noise model is Middleton Class-A. The \ac{pdf} of a real-valued noise sample \(n_i\) is an infinite mixture of Gaussian distributions with the mixture weights following a Poisson distribution \cite{oh:2017, lin:2013, shongwey:2014}:
\begin{equation}
    \label{eq:impulsive-noise}
    f(n_i)=\sum_{m=0}^{\infty}\frac{A^{m} \exp(-A)}{m!}\,\frac{1}{\sigma_m\sqrt{2\pi}}
    \exp\!\Big(-\frac{n_i^{2}}{2\sigma_m^{2}}\Big),
\end{equation}
where \(m\sim\mathrm{Poisson}(A)\), \(n_i|m\sim\mathcal{N}(0,\sigma_m^2)\) with \(\sigma_m^{2}=\sigma_g^{2}\left(\frac{m}{A\Gamma} + 1\right), 
\Gamma=\frac{\sigma_g^{2}}{\sigma_I^{2}}\). Here \(\sigma_g^{2}\) is the background Gaussian noise power, \(\sigma_I^{2}\) is the impulsive-noise power, and the mean total power is \(\sigma_{\text{total}}^{2}=\sigma_g^{2}+\sigma_I^{2}\). For complex baseband, apply the real-valued model independently to \(I/Q\), yielding a circularly symmetric complex process.
\section{Methodology}
\label{sec:methodology}
\begin{figure}[!hbpt]
    \centering
    \includegraphics[width=\linewidth]{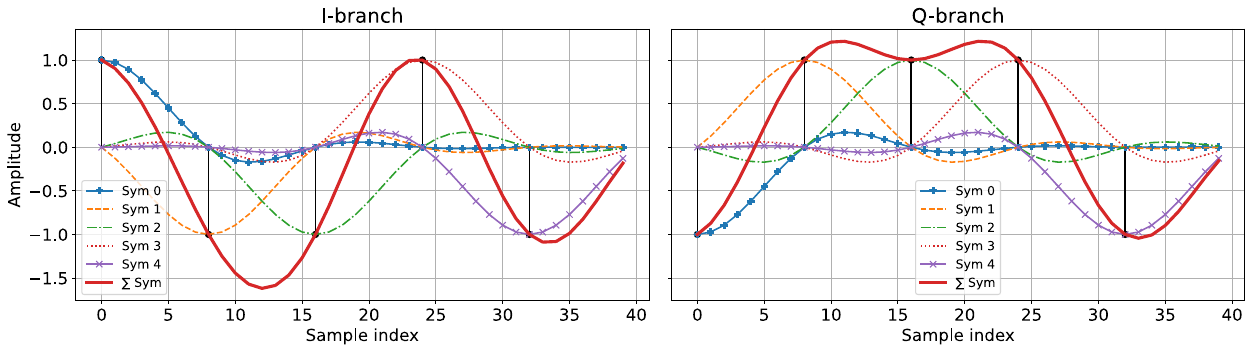}
    \caption{Visualization of \ac{isc} introduced by pulse shaping. Each colored segment represents the contribution of a distinct symbol to the overall waveform. Due to pulse overlap, each sample contains information from multiple adjacent symbols, creating structured context that can be exploited.}
    \label{fig:isi_viz}
\end{figure}

This work investigates how the deterministic structure introduced by pulse shaping can be exploited to learn meaningful representations of communication signals. In particular, we hypothesize that the \ac{isc} arising from pulse shaping introduces a form of contextual dependency between samples that can be effectively modeled using attention-based architectures. Fig.~\ref{fig:isi_viz} illustrates how overlapping symbol contributions create such structured sample-level context.

\begin{figure}[!hbpt]
    \centering 
    \includegraphics[width=0.85\linewidth]{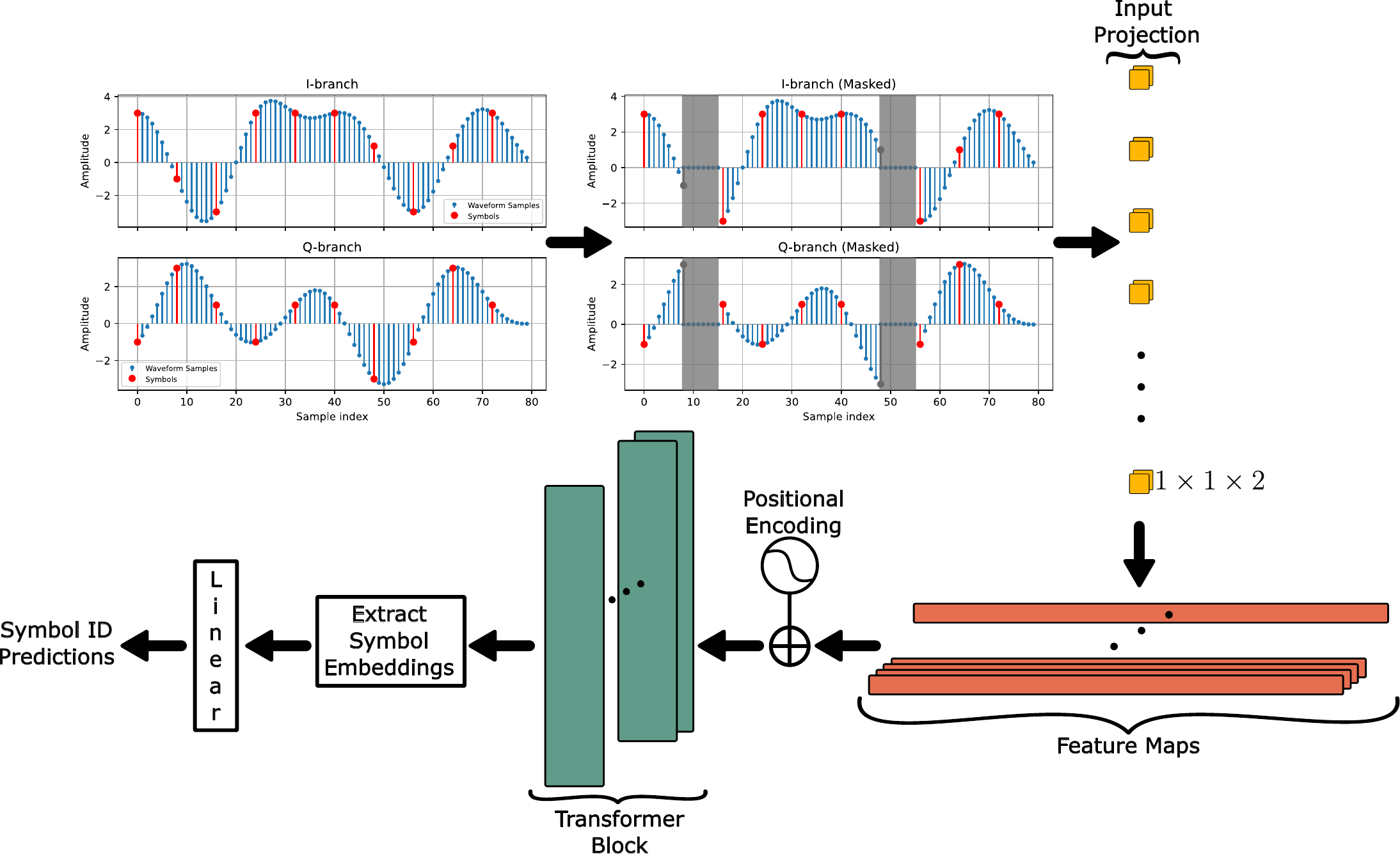}
    \caption{Conceptual diagram of end-to-end \ac{msm} architecture.}
    \label{fig:model-architecture}
\end{figure}

We adopt a \ac{bert}-style training framework with a Transformer \ac{nn}. The model input is a sequence of complex-valued baseband symbols, which are mapped to oversampled time-domain waveforms using pulse shaping filters. During training, \(15\%\) of the symbols are randomly masked by setting their corresponding sample spans to zero. A discrete vocabulary assigns a unique symbol \ac{id} to each distinct constellation point across all considered modulations. The model is trained to predict the correct \ac{id} for each masked symbol using only surrounding unmasked samples, thereby internalizing the contextual structure imposed by pulse shaping and \ac{isc} (Fig.~\ref{fig:model-architecture}).

While the primary objective is to study the contextual properties of pulse-shaped signals, we additionally demonstrate the utility of the model by applying it to recover symbols corrupted by impulsive noise.
\subsection{Training and Inference Signals}
\label{subsec:signal-model}
For training, the model uses the non-impaired basebands signal \(s[n]\) as defined in Sec.~\ref{subsec:signal-model}. At inference, we evaluate the model on signals corrupted by additive Middleton Class-A noise, \(z[n]\), resulting in the received signal:
\begin{equation}
    \label{eq:noisy-signal}
    y[n] = s[n] + z[n], \quad z[n] \sim \text{Middleton Class-A}(A, \Gamma).
\end{equation}
\begin{table}[!hbpt]
\caption{Dataset description.}
\label{table:dataset-description}
\centering
\renewcommand{\arraystretch}{1.2}
\setlength{\tabcolsep}{8pt}
\begin{tabular}{ll}
\toprule
\textbf{Description} & \textbf{Range} \\
\midrule
Modulation types & \{BPSK, QPSK, PSK8, PSK16, QAM4, QAM16, QAM64, QAM256\} \\
Symbol rate (symbols/s) & 1 \\
\Ac{sps} & 8 \\
Filter span (symbols) & \{10, 12, 14, 16\} \\
Roll-off factor & \{0.25, 0.35, 0.45, 0.55, 0.65, 0.75\} \\
\bottomrule
\end{tabular}
\end{table}
\subsection{Data Generation Pipeline}
\label{subsec:data-gen-pipeline}
Waveforms are generated on-the-fly using a modular NumPy/PyTorch pipeline. For each example, a modulation scheme is sampled, symbols are drawn uniformly from the corresponding constellation, and mapped to \(I/Q\) components. These symbols are represented as two real-valued arrays (\(I\) and \(Q\)) rather than a single complex array. Each branch is pulse-shaped with a \ac{rc} \ac{fir} filter, with span and roll-off uniformly sampled from Table~\ref{table:dataset-description}. Waveforms are normalized to unit power, with \ac{sps} fixed at \(8\) to ensure \(1024\)-sample inputs for the Transformer. Each waveform is paired with a target sequence corresponding to correct symbol \acp{id} for supervised training. These symbol sequences serve as the ground-truth labels for masked positions during training. Training uses only non-impaired waveforms. At inference, impulsive noise can be optionally added using a Middleton Class-A noise model in a fully vectorized implementation. Code is available at the following repository: \url{https://github.com/OguzBedir/Masked_Symbol_Modeling.git}.
\subsection{System Model}
Each training waveform passes through a masking module that zeros the samples of a random \(15\%\) of symbols. The masked signal is processed by the \texttt{Masked Symbol Transformer}, which maps the two input channels to a \(512-\)dimensional embedding via a learnable \(1\)D projection layer, adds fixed sinusoidal positional encoding, and applies six Reformer blocks with \ac{lsh} attention (bucket size 64, four hashes), shared weights, and reversible layers for memory efficiency. For each masked symbol, embeddings over its sample span are mean-pooled and passed to a linear classifier mapping \(\mathbb{R}^{512} \to \mathbb{R}^{V}\), where \(V=272\) is the vocabulary size. Loss is computed with cross-entropy only over masked symbols, with inverse-frequency weighting to mitigate class imbalance.
\subsection{Training Setup}
Training is fully self-supervised, using an on-the-fly \texttt{IterableDataset}. No external datasets or pre-recorded signals are used. Optimization uses Adam (\(10^{-3}\) learning rate) on all parameters jointly. Experiments are run on a single NVIDIA A100 for \(24\) hours, corresponding to \(37{,}551\) training steps, with a batch size of \(64\). As data are generated procedurally, there is no notion of epochs.
\section{Experimental Setup \& Results}
\label{sec:experimental-setup-results}
We evaluate \ac{msm} for predicting symbols affected by impulsive noise, demonstrating its utility in a representative application. All experiments use the same base waveform configuration as in training: each waveform has \(8\) \ac{sps}, a filter span (in symbols) and a roll-off factor uniformly sampled from the corresponding values listed in Table~\ref{table:dataset-description}. Each waveform consists of \(1024\) samples, and is trimmed to remove leading and trailing transients introduced by filter delays, ensuring alignment between pulse-shaped waveforms and symbol boundaries, identical to the training configuration. The embedding dimension is fixed at \(512\). To ensure statistical reliability and avoid favorable outcomes due to random initialization, all experiments in this section are repeated with three different random seeds, and the average performance is reported. For each seed, the simulation is run on at least \(10{,}000\) target symbols.

\begin{figure}[!hbpt]
    \centering
    \includegraphics[width=0.6\linewidth]{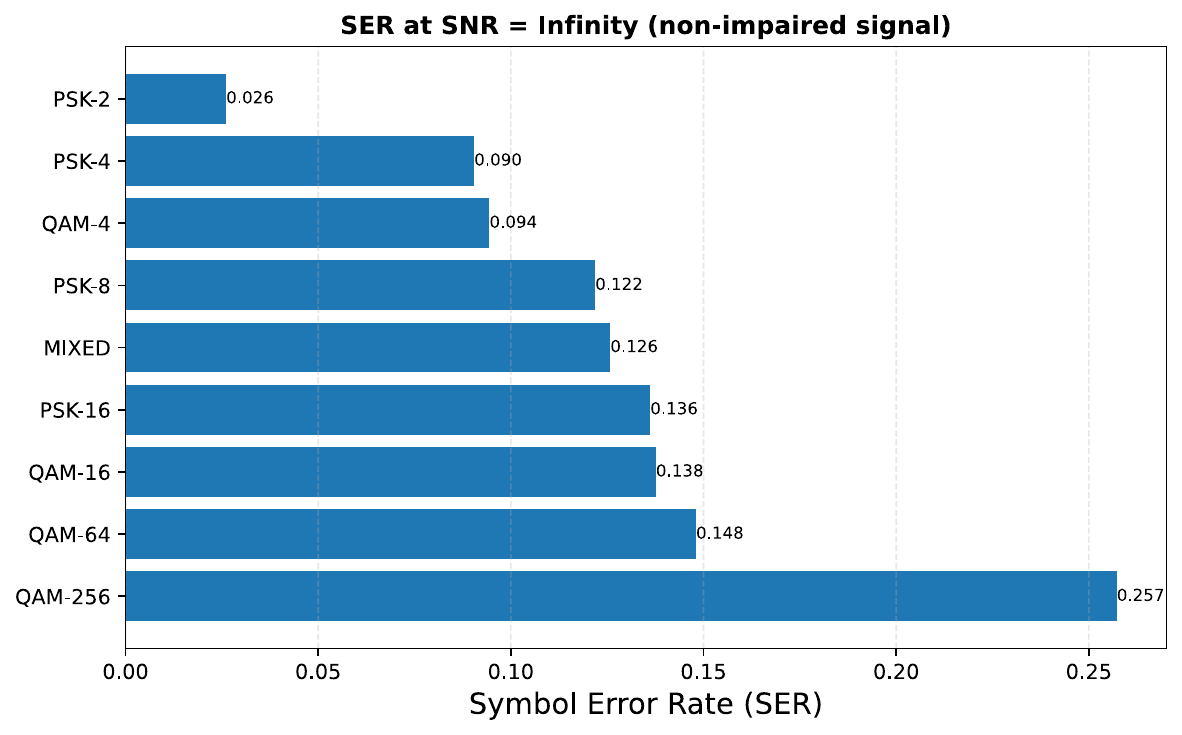}
    \caption{\Acf{ser} for different modulation schemes at infinite \acf{snr} (i.e., non-impaired signals).}
    \label{fig:m15-none-clean}
\end{figure}

We first evaluate \ac{msm} on non-impaired signals, under conditions identical to those used in training: \(15\%\) of the symbols per waveform are randomly masked, and the model predicts the corresponding symbol \acp{id}. Results are reported separately for each modulation scheme, as well as for a “mixed” setting in which the modulation scheme is sampled uniformly at random from Table~\ref{table:dataset-description}. Each waveform contains \(128\) symbols, and inference is performed in batches of \(64\) waveforms. With \(15\%\) masking, this yields \(19\) masked symbols per waveform, or \(1{,}216\) prediction targets per batch. Nine batches are processed per seed, resulting in \(10{,}944\) predictions. As expected, the model achieves higher accuracy on simpler modulation formats, with performance remaining stable across most schemes as shown in Fig.~\ref{fig:m15-none-clean}. Notable deviations occur for \ac{bpsk} and \ac{qam}-\(256\), where accuracy is significantly higher or lower, respectively, due to the extreme simplicity or complexity of their constellations.

To assess utility under impairments, we evaluate the model on waveforms impaired by Middleton Class-A impulsive noise. The fully corrupted waveform is not fed directly to the model; instead, symbols affected by impulsive noise are identified, and their corresponding waveform segments are masked, as illustrated in Fig.~\ref{fig:model-architecture}, while unaffected segments are left unchanged. The model is then tasked with recovering the symbol \acp{id} at these masked positions using contextual information from surrounding unmasked samples.

\begin{figure}[!hbpt]
  \centering
  \begin{subfigure}{0.49\linewidth}
    \includegraphics[width=\linewidth]{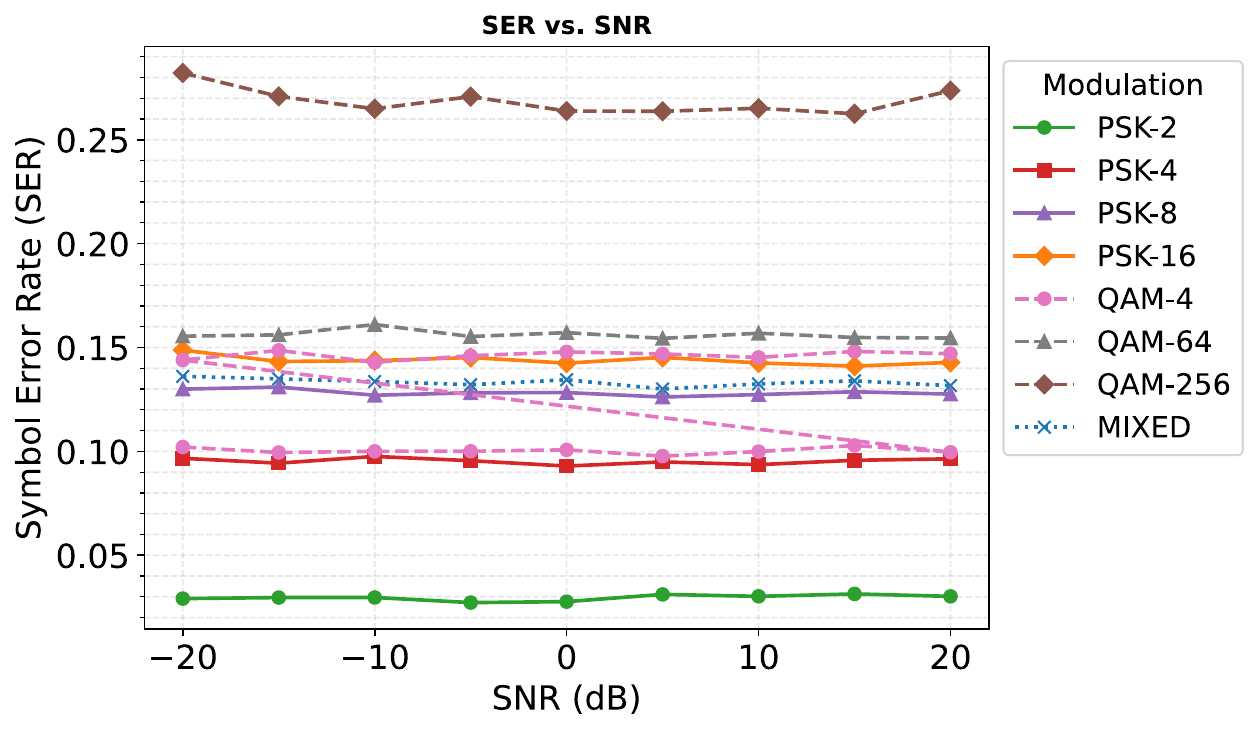}
    \caption{\acs{ser} under strong impulsive noise (\(\Gamma=10^{-6}\)).}
    \label{subfig:custom-middleton-a1-g1}
  \end{subfigure}\hfill
  \begin{subfigure}{0.49\linewidth}
    \includegraphics[width=\linewidth]{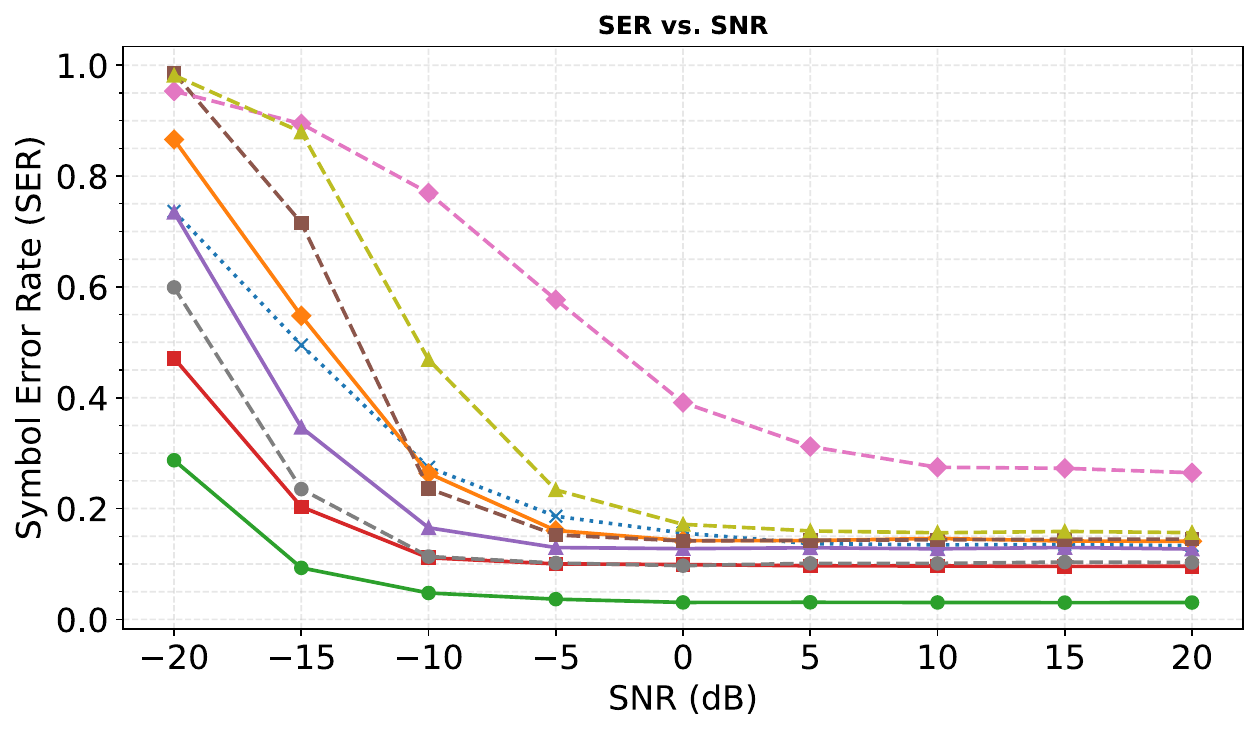}
    \caption{\acs{ser} under moderate impulsive noise (\(\Gamma=10^{-3}\)).}
    \label{subfig:custom-middleton-a1-g2}
  \end{subfigure}
  \caption{\Acf{ser} of the proposed model under Middleton Class-A noise for two \(\Gamma\) values with \(A\) set for an average \(15\%\) symbol-hit rate.}
  \label{fig:custom-middleton-a1}
\end{figure}

This evaluation represents a semi-synthetic impairment scenario, designed to isolate the model’s ability to exploit contextual structure rather than to assess robustness against widespread noise. We use \(\Gamma \in \{10^{-3}, 10^{-6}\}\) \cite{lin:2013} to ensure that the Gaussian background noise component is insignificant and that the impairment is dominated by impulsive noise. For both cases in Fig.~\ref{fig:custom-middleton-a1}, the impulsive index is set to \(A = \frac{-\ln(0.85)}{L}\big|_{L=8}\) corresponding to an average symbol-hit rate of \(15\%\).

Since the noise generation process is modeled faithfully according to the Middleton Class-A distribution, it remains inherently stochastic. Consequently, achieving exactly \(15\%\) symbol hits is not possible unless (i) high-cost rejection sampling is applied, or (ii) a direct symbol-selection approach is used, which would violate proper statistical noise modeling. The derivation of the chosen \(A\) value and the computation of its confidence interval are provided in Appendix~\ref{appx:imp-index-calc}.

In the configuration shown in Fig.~\ref{subfig:custom-middleton-a1-g1}, the Gaussian-to-impulsive noise ratio \(\Gamma\) is so small that the Gaussian component is effectively negligible; and impulsive bursts dominate the \ac{snr}. Because the masking removes affected symbols and Gaussian noise power is minimal, performance remains nearly constant across \ac{snr} values.

The behavior differs in Fig.~\ref{subfig:custom-middleton-a1-g2}, where \(\Gamma\) is larger and the Gaussian component is no longer negligible. In the low-\ac{snr} regime, Gaussian noise significantly degrades performance, even after masking impulsive noise–affected symbols. As the \ac{snr} increases, the influence of the Gaussian component diminishes, and performance stabilizes; similar to that observed in the first configuration.
\section{Conclusion}
\label{sec:conclusion}
This work investigates how the deterministic contextual structure introduced by pulse shaping can be exploited for symbol inference. We demonstrate that Transformer networks, through their attention mechanism, can leverage this structured context to predict symbol \acp{id} from surrounding “in-between” samples.
The results support our central hypothesis that it is possible to design receivers that go beyond conventional symbol detection and instead interpret the transmitted waveform structure to recover information. This capability may enhance error correction and other downstream tasks. While promising, the findings are preliminary and require systematic ablation to fully understand trade-offs and limitations.

A key direction for improvement is refining the input representation. In our current setup, the model processes signals in their native \(\mathbb{R}^2\) form, with separate \(I\) and \(Q\) channels. An alternative is to quantize the amplitudes and define the vocabulary over quantization-level pairs \((I, Q)\), assigning a distinct embedding vector to each pair, including a dedicated “mask” embedding. All embeddings, including the mask, would be initialized randomly and learned during training, enabling the model to develop richer, semantically meaningful representations. When trained across diverse signal and channel conditions, these embeddings could encode both intrinsic waveform properties and channel characteristics. Such changes would make the framework more faithful to the original BERT formulation and could remove the need for the current input projection layer.

Learning the mask embedding would encourage a more expressive latent space, enabling operation under more challenging conditions. Explicit masking at inference, used in our current setup, may then be unnecessary, as the model could process unmasked waveforms directly while retaining performance. Another direction is shifting from symbol-level to sample-level prediction, potentially improving contextual modeling and robustness.

Future ablations should examine: (i) embedding dimension, (ii) Transformer depth, attention heads, and encoder/decoder variants, and (iii) masking strategies, including percentage and learned versus fixed mask embeddings, and (iv) a comparison with alternative deep learning architectures that do not use an attention mechanism to determine whether attention provides substantial gains in leveraging signal context. These studies will clarify how representation, architecture, and masking interact, guiding broader applicability to diverse and challenging channels.
\begin{ack}
This material is based upon work supported by the U.S. Department of Energy, Office of Science, Office of Advanced Scientific Computing Research, Early Career Research Program under Award Number DE-SC0023957.

Portions of this research were conducted with the advanced computing resources provided by Texas A\&M High Performance Research Computing.
\end{ack}
\bibliographystyle{unsrtnat}
\bibliography{main}
\newpage
\appendix
\section{Impulsive Index: Calibration and Concentration}
\subsection{Calibrating the Impulsive Index}
\label{appx:imp-index-calc}
Let \(\{m[n]\}_{n=0}^{N-1}\) denote the per-sample impulsive counts in the Middleton Class-A model, with \(m[n]\overset{\text{i.i.d.}}{\sim}\mathrm{Poisson}(A)\). A sample is affected by impulsive noise iff \(m[n]>0\), which occurs with probability
\begin{equation}
    p_{\text{sample}} \triangleq \mathbb{P}(m[n]>0)=1-e^{-A}.   
\end{equation}

Let \(K\triangleq N/L\) be the number of symbols in a waveform. For each symbol index \(i\in\{0,\dots,K-1\}\), define the sample index set \(\mathcal{I}_i\triangleq\{i \cdot L, i \cdot L+1, \dots,(i+1) \cdot L-1\}\). A symbol is affected iff at least one of its \(L\) samples is affected. By independence across samples,
\begin{align}
    p_{\text{sym}} &\triangleq \mathbb{P}(\mathds{1}\{\exists n \in \mathcal{I}_{i}: m[n] > 0\} = 1) = 1 - \mathbb{P}(m[n] = 0, \forall n \in \mathcal{I}_{i}) \\
    &= 1 - (1 - p_{\text{sample}})^{L} = 1 - e^{-A \cdot L}. \nonumber
\end{align}
Since \(\mathds{1}\{\exists n \in \mathcal{I}_{i}: m[n] > 0\}\) are \ac{iid} across disjoint symbols, the number of affected symbols
\begin{equation}
    S \triangleq \sum_{i=0}^{K-1} \mathds{1}\{\exists n \in \mathcal{I}_{i}: m[n] > 0\} \sim \mathrm{Binomial}\!\big(K,\,p_{\text{sym}}\big),
\end{equation}
which yields \(\mathbb{E}\!\left[\tfrac{S}{K}\right]=p_{\text{sym}}\).

\paragraph{Targeting a desired symbol-hit rate.}
To target an \emph{average} fraction \(p^{\star}=0.15\) of affected symbols per waveform, set \(p_{\text{sym}}=p^{\star}\) and solve for \(A\):
\begin{equation}
     A^\star = -\frac{1}{L}\,\ln(1-p^{\star})=\frac{-\ln(0.85)}{L}.   
\end{equation}

This calibration ensures \(\mathbb{E}[S/K]=p^{\star}\). Because \(S\) is binomial, the realized fraction \(\hat p=S/K\) fluctuates around \(p^{\star}\) for finite \(K\); achieving \emph{exactly} \(15\%\) per waveform is impossible without additional (and statistically distorting) conditioning such as rejection sampling or direct symbol selection.

\subsection{Concentration of the Empirical Affected-Symbol Ratio}
Let \(p\triangleq p_{\text{sym}}\). For any \(\epsilon\in\big(0,\min\{p,1-p\}\big)\), a standard two-sided Chernoff bound for Bernoulli means \cite{impagliazzo:2010} gives
\begin{equation}
    \mathbb{P}\left(\lvert \hat p - p\rvert \ge \epsilon\right)\le 2\exp\Big(-K D_{\mathrm{KL}}(p+\epsilon\|p)\Big),
\end{equation}
where, for \(q,p\in(0,1)\), the binary \ac{kl} divergence is
\begin{equation}
    D_{\mathrm{KL}}(q\,\|\,p) = q\ln\frac{q}{p} + (1-q)\ln\frac{1-q}{1-p}.   
\end{equation}
There is no closed-form inverse for \(\epsilon\) as a function of the tail probability \(\delta\), but for small \(\epsilon\) a quadratic expansion yields \cite{boucheron:2004}
\begin{equation}
    D_{\mathrm{KL}}(p+\epsilon\,\|\,p) \approx \frac{\epsilon^2}{2\,p(1-p)}
    \Longrightarrow
    \epsilon \approx \sqrt{\frac{2\,p(1-p)}{K}\,\ln\Big(\tfrac{2}{\delta}\Big)}.   
\end{equation}
With \(\delta=0.05\), \(p=0.15\), and \(K=128\), this gives \(\epsilon \approx 0.085726\), i.e.,
\begin{equation}
    \hat p \in \big[p-\epsilon,\; p+\epsilon\big] \approx [0.064274,0.235726],
\end{equation}
with probability at least \(95\%\).
\end{document}